\documentclass[aps,showpacs,twocolumn]{revtex4}
\usepackage{epsfig}
\usepackage{color}
\usepackage{amsmath}
\usepackage{amssymb}
\usepackage{bbm}

\newcommand{\be}{\begin{equation}}
\newcommand{\ee}{\end{equation}}
\newcommand{\bea}{\begin{eqnarray}}
\newcommand{\eea}{\end{eqnarray}}

\begin{document}

\title{Magnon-induced superconductivity in field-cooled spin-1/2 antiferromagnets}

\author{Naoum Karchev}

\affiliation{Department of Physics, University of Sofia, 1164 Sofia, Bulgaria}

\begin{abstract}

If, during the preparation, an external magnetic field is applied upon cooling we say it has been field cooled.
A novel mechanism for insulator-metal transition and superconductivity in field-cooled spin-$1/2$ antiferromagnets on bcc lattice is discussed. Applying  a magnetic field along the sublattice B magnetization, we change the magnetic and transport properties of the material. There is a critical value $H_{cr1}$. When the magnetic field is below the critical one $H<H_{cr1}$ the prepared material is a spin$-1/2$ antiferromagnetic insulator. When $H>H_{cr1}$ the sublattice A electrons are delocalized and the material is metal. There is a second critical value $H_{cr2}>H_{cr1}$. When $H=H_{cr2}$, it is shown that the Zeeman splitting of the  sublattice A electrons is zero and they do not contribute to the magnetization of the system. At this quantum partial order point (QPOP) the sublattice B transversal spin fluctuations (magnons) interact with sublattice A electrons inducing spin anti-parallel \emph{p}-wave superconductivity which coexists with magnetism. At zero temperature the magnetic moment of sublattice B electrons is maximal. Below the N\'{e}el temperature $(T_N)$ the gap is approximately constant with a small increase when the system approaches $T_N$. It abruptly falls down to zero at temperatures above  $T_N$.

\end{abstract}

\pacs{75.50.Ee,74.20.Mn,74.20.Rp}

\maketitle

\section {Introduction}
The preparation of a superconductor from magnetically ordered material is an old-standing problem in solid state physics. The solution of this problem arrived at  great success when J. G. Bednorz and K. A. M\"{u}ller  prepared a high-temperature superconductor from an antiferromagnet\cite{Bednorz86}. The parent material LaCuO$_4$ is an antiferromagnetic insulator. The replacement of lanthanide with barium La$_{1-x}$Ba$_x$CuO$_4$ suppresses the antiferromagnetic order in favor of superconductivity.
The articles \cite{Vojta09,Taillefer10} overview the current understanding of the cuprate high-temperature superconductivity.

The parent compound of the Fe-based superconductor LaOFeAs is an antiferromagnetic metal. Replacing the oxygen with phosphorus LaO$_{1-x}$F$_x$FeAs one obtains a superconductor with $T_{sc}=26 K$ \cite{Fe}. Another way to prepare the Fe-based superconductor is by
replacement of one pnictide element $As$ by another $P$ \cite{As-P}

 With distinction to cuprate superconductors the Fe-based superconductors can be induced by pressure \cite{Fe-pressure}. Under the pressure, the N\'{e}el temperature of SrFe$_2$As$_2$ decreases and at $P_c=3.6 GPa$ it becomes equal to zero. The pressure-induced superconductivity abruptly emerges close to the phase boundary at $P_c$. Works on Fe-based superconductors are reviewed, for example, in \cite{Wen12,Chubukov15}.

The typical temperature-pressure phase diagrams observed in the heavy fermion materials show that at ambient pressure the compounds order into antiferromagnets below the N\'{e}el temperature $T_N$. Applying pressure
reduces $T_N$ monotonically. The QCP is the critical pressure $P_c$ at which the N\'{e}el temperature $T_N = 0$. In CePd$_2$Si$_2$ the QCP, is not explicitly observed. Instead, a superconductivity emerges at very
low temperatures in the vicinity of $P_c$\cite{CePdSi,Mathur98,Kenzelmann08}. The theory of the spin fluctuation mechanism of superconductivity is successful in explaining the superconductivity around the antiferromagnetic quantum critical point in heavy electron systems \cite{Miyake86,Emery86}.

The coexistence of ferromagnetism and superconductivity as cooperative phenomena was first discovered in UGe$_2$ under pressure \cite{UGe2}. At ambient pressure UGe$_2$ is an itinerant ferromagnet below the Curie temperature. With increasing pressure the system passes through two successive quantum phase transitions, from ferromagnetism to FM-superconductivity at $P\sim 10 kbar$, and at higher pressure $Pc\sim 16 kbar$ to paramagnetism.

Another important U-based superconductor is URhGe. It is strongly anisotropic ferromagnet below the Curie temperature of 9.5K with a magnetic moment directed along the c-axis. For temperatures below $260 mK$ superconductivity was observed \cite{HH}. At low enough temperature the magnetic field, about 1.3 Tesla, directed along the b-axis suppresses the superconducting state but at much higher field, about 8 Tesla, the superconductivity is recreated and exists till the field, about 13 Tesla \cite{Levy}.

If, during the preparation, an external magnetic field is applied upon cooling we say it has been field cooled.
We discuss a novel mechanism for insulator-metal transition and superconductivity in field-cooled spin-$1/2$ antiferromagnets on bcc lattice. Applying  a magnetic field on sublattice A and B electrons along the sublattice B magnetization, during preparation of the material, we change the magnetic and transport properties of the material. There is a critical value $H_{cr1}$. When the magnetic field is below the critical one $H<H_{cr1}$ the prepared material, with switched off magnetic field, is spin$-1/2$ antiferromagnetic insulator, while when $H>H_{cr1}$ the sublattice A electrons are delocalized and the material is metal, with well defined  Fermi surfaces and Fermi liquid-type quasiparticles. This method of preparation of ferrimagnetic spinel is very popular \cite{spinel+,spinelFeCr2S4,spinelCv1,spinel08,spinel++,spinel11b,spinel11a,spinel11c,spinel12a,spinel12b,spinel+1,spinel+2,spinel+3,FeV2O4-14}. The magnetization-temperature and magnetic susceptibility curves for zero-field-cooled (ZFC) and field-cooled (FC) ferrimagnetic spinel display a notable difference below N\'{e}el $T_N$ temperature.

The magnetic properties of the system under consideration in the present paper can be described with an effective model of a two-sublattice spin system with $s^B=1/2$ and the sublattice A spin $s^A<1/2$ is a varying parameter which accounts for the magnetic field \cite{Karchev15a}. It decreases when the applied field increases.

There is a second critical value $H_{cr2}>H_{cr1}$. When $H=H_{cr2}$, it is shown that the Zeeman splitting of the  sublattice A electrons is zero and they do not contribute to the magnetization of the system. At this quantum partial order point (QPOP) the sublattice B transversal spin fluctuations (magnons) interact with sublattice A electrons inducing spin anti-parallel p-wave superconductivity which coexists with magnetism. At zero temperature the magnetic moment of sublattice B electrons is maximal.

There are many papers devoted to superconductivity induced by spin fluctuations. Some of them investigate itinerant systems in which the spin-$1/2$ fermions responsible for the magnetism are the same quasiparticles which form the Cooper pairs \cite{EnzMatthias,FayApple80,Kirkpatrick01,Roussev01,Karchev03}.

Spin-fermion model describes materials which get their magnetic properties from a system of localized magnetic moments being coupled to conducting electrons \cite{cfm01,cfm02,cfm021,cfm03,cfm04,cfm05,cfm06}.

The exchange of itinerant and localized electrons leads to Zeeman splitting of the delocalized electrons, which in turn, suppresses the magnon induced superconductivity.

In the present paper we consider one band model of field-cooled antiferromagnetic. When a magnetic field is applied along the sublattice B magnetization, the electrons are confined spatiali in a different way. Sublattice B electrons are localized, while sublattice A ones are delocalized. Different electrons form Cooper pairs and magnetic moment, but all these electrons are from one band. The effective model is spin-fermion model with Zeeman splitting of itinerant electrons compensated by the applied field. This reminds us of the Jaccarino- Peter (JP) compensation mechanism\cite{JP62}. In a rare earth ferromagnetic metal the conduction electrons are in an effective field due to the exchange interaction with the rare earth spins. It is in general so large as to inhibit the occurrence of superconductivity. For some systems the exchange interaction have a negative sign. This  allows for the conduction electron polarization to be canceled by an external magnetic field so that if, in addition these metals possess phonon-induced attractive electron-electron interaction, superconductivity  occurs in the compensation region. If the effective field is not large the coexistence of superconductivity end magnetic order is possible and the external magnetic field enhances the superconductivity. This is not enough in strongly correlated systems because the Coulomb repulsion localizes the electrons which inhibits the superconductivity. In the present paper we show that applied magnetic field delocalizes the sublattice A electrons which permits the formation of Cooper pairs.

\section {Novel mechanism of insulator-metal transition}

To begin with we consider a theory with the Hamiltonian
\bea \label{antiferro1}\nonumber
& h &  =-t\sum\limits_{\ll ij \gg _A } {\left( {c_{i\sigma }^ + c_{j\sigma } + h.c.} \right)}
  +U\sum\limits_{i\in A} n_{i\uparrow}n_{i\downarrow} \\
& - & \mu \sum\limits_{i\in A} {n_i}
 -  J_A\sum\limits_{  \ll  ij  \gg_A  } {{\bf S_i^A}\cdot {\bf S_j^A}}  \\
 & + & J\sum\limits_{  \langle  ij  \rangle } {{\bf S_i^A}}\cdot {\bf S_j^B}
  -  J_B\sum\limits_{  \ll  ij  \gg_B  } {{\bf S_i^B}
\cdot {\bf S_j^B}}, \nonumber
\eea
where $t>0$ is the hopping parameter,  ${\bf S_i^A}$ is the spin of the itinerant
electrons at the sublattice $A$ site with components $S^{\nu A}_i=\frac 12\sum\limits_{\sigma\sigma'}c^+_{i\sigma}\tau^{\nu}_{\sigma\sigma'}c^{\phantom +}_{i\sigma'}$,  $(\tau^x,\tau^y,\tau^z)$ are the Pauli matrices, ${\bf S_i^B}$ is the spin of the localized electrons  at the sublattice $B$ site, $\mu$
is the chemical potential, $n_{i\sigma}=c^+_{i\sigma}c_{i\sigma}$ and $n_i=n_{i\uparrow}+n_{i\downarrow}$. The
sums are over all sites of a body centered cubic lattice, $\langle i,j\rangle$ denotes the sum over the nearest neighbors, while $ \ll  ij  \gg_A$ and  $\ll  ij  \gg_B$ are sums over all sites of sublattice $A$ and $B$ respectively. The Heisenberg terms describe ferromagnetic Heisenberg exchange between sublattice A $(J_A>0)$ and sublattice B $(J_B>0)$ electrons, while the term  $J>0$ is the antiferromagnetic exchange constant between localized and itinerant electrons. The term with the constant $U>0$ is the Coulomb repulsion.

We represent the Fermi operators, the spin of the itinerant electrons and the density operators of sublattice A electrons in terms of the Schwinger-bosons ($\varphi_{i,\sigma}, \varphi_{i,\sigma}^+$) and slave fermions ($h_i, h_i^+,d_i,d_i^+$) (\ref{QCB2}).
An important advantage of working with Schwinger bosons and slave fermions
is the fact that Hubbard term is in a diagonal form. The fermion-fermion and fermion-boson interactions are included in the hopping term. One treats them as a perturbation with parameter $t/U$. To proceed we keep only the quadratic, with respect to fermions, terms. This means that the averaging in the subspace of the fermions is performed in one fermion-loop approximation. The other terms in $t/U$ expansion are dropped. In the present paper we consider antiferromagnetic insulator, hence the Coulomb repulsion $U$ is much larger then the hopping parameter $t/U<<1$. Therefore, our approximation is very appropriate.
We use the Holstein-Primakoff representation of the spin operators of sublattice B localized electrons ${\bf S^B_j}(a^+_j,a_j)$, where $a^+_j,\,a_j$
are Bose fields (\ref {supp9}), (\ref {supp9a}), (\ref {supp10}). We account for the effect of the applied, during the preparation, magnetic field on itinerant electrons adding the term $-H \sum\limits_{i\in A} {S^{zA}_{i}}$ into the Hamiltonian (\ref{antiferro1}). The magnetic field is applied along the direction set by the magnetization of the sublattice B localized spins.
Then the Hamiltonian for the free $d$ and $h$ Fermions reads (see Appendix A)
\be\label{antiferro2}
h_0 =   \sum\limits_{k\in B_r} \left (\varepsilon^d_k d_k^+ d_k + \varepsilon^h_k h_k^+ h_k \right)\ee
with dispersions
\bea\label{antiferro3}
\varepsilon^d_k & = & -4t\varepsilon_k +U-\mu+2J-\frac H2 \nonumber \\
\varepsilon^h_k & = & 4t\varepsilon_k+\mu+2J -\frac H2  \\
\varepsilon_k &  = & \left ( \cos k_x+\cos k_y+\cos k_z \right)\nonumber \eea
The ground state of the system, with free-fermion Hamiltonian (\ref{antiferro2}) is labeled by the density of electrons
\be\label{QCB4d} n=1-<h^+_i h_i>+<d^+_id_i> \ee
 and the zero temperature spontaneous magnetization of the electron
\begin{equation}
m=\frac 12 \left(1-<h^+_i h_i>-<d^+_id_i>\right). \label{QCB5}
\end{equation}
At half-filling
\be\label{QCB4e} <h^+_i h_i>=<d^+_id_i>. \ee
To solve this equation, for all values of the parameters, one sets the chemical potential $\mu=U/2$.
Utilizing this representation of $\mu$ we calculate the dispersions of $"d"$ and $"h"$ fermions (\ref{antiferro3}) as a function of the applied magnetic field. To this end it is convenient to introduce the critical magnetic field
\be\label{antiferro4}
H_{cr1}=U+4J-24t
\ee
Having in mind equations (\ref{antiferro3}), with $\mu=U/2$, and the equation (\ref{antiferro4}) the Fermion dispersions  $\varepsilon^d_k$ and $\varepsilon^h_k$
adopt the form
\bea\label{antiferro4a}
\varepsilon^d_k & = & 4t\left[-\varepsilon_k +3 +\frac {H_{cr1}-H}{8t} \right] \nonumber \\
\varepsilon^h_k & = &  4t\left[\varepsilon_k +3 +\frac {H_{cr1}-H}{8t} \right]
\eea
 When the applied magnetic field is below the critical one $H<H_{cr1}$ the Fermion dispersions are positive ($\varepsilon^d_k>0$, \,\,$\varepsilon^h_k>0$)  for all values of the wave vector $k$. The minimum of the $d$-fermion dispersion is at the center of the Brillouin zone of a cubic lattice $B_r$ $\textbf{k}=(0,0,0)$ and $\varepsilon^d_0=(H_{cr1}-H)/2$. The minimum of the  $h$-fermion dispersion is at the vertexes of the Brillouin zone $\textbf{k}^*=(\pm \pi,\pm \pi,\pm \pi)$ and $\varepsilon^h_{k^*}=(H_{cr1}-H)/2$.
 This means that Fermions excitations are gapped which is our definition for insulating state. If one applies magnetic field below the critical one the prepared material is insulator.

When the applied field is above the critical one $H>H_{cr1}$, the solutions of the equations
 \bea\label{antiferro4b}
\varepsilon^d_k & = & 4t\left[-\varepsilon_k +3 +\frac {H_{cr1}-H}{8t} \right]=0 \nonumber \\
\varepsilon^h_k & = &  4t\left[\varepsilon_k +3 +\frac {H_{cr1}-H}{8t} \right]=0
\eea
define the Fermi surfaces of $"d"$ and $"h"$ quasiparticles.
The resultant material is metal. The system possesses a novel insulator-metal transition when magnetic field is applied and the critical value is $H_{cr1}$ (\ref{antiferro4}).

The equation (\ref{antiferro4a}) shows that when the magnetic field is zero the system is insulator if Coulomb repulsion is strong. When a hydrostatic pressure is applied the hopping parameters $t$ increases, and for $24t>U+4J$ the system is metal. The point is that under a hydrostatic pressure all electrons in the material delocalize, while when a magnetic field is applied the electrons in the system are geometrically separated and sublattice A electrons are delocalized, but sublattice B ones are localized. This is important novelty.

There is a second critical value
\be\label{antiferro41}
H_{cr2}=U+4J. \ee
When $H=H_{cr2}$ the material is metal ($H_{cr2}>H_{cr1}$) and Zeeman splitting of sublattice A electrons is zero. The Fermion dispersions  $\varepsilon^d_k$ and $\varepsilon^h_k$ (\ref{antiferro4a}) adopt the form
\bea\label{antiferro4c}
\varepsilon^d_k & = & -4t\varepsilon_k  \nonumber \\
\varepsilon^h_k & = &  4t\varepsilon_k
\eea
With dispersions (\ref{antiferro4c}) spontaneous magnetization $m$ (\ref{QCB5}) of sublattice A electrons is zero and they do not contribute the magnetization of the system. At this critical point the system is in partial order state. Only sublattice B electrons are magnetically ordered, while sublattice A electrons are magnetically disordered.

Partial order is well known phenomenon and has been subject of extensive studies. The partial order has been predicted in frustrated antiferromagnetic systems\cite{Diep97} and ferrimagnets \cite{Karchev08b,Karchev09b}. An effective description of bipartite antiferromagnets with magnetically uncompensated sublattices. Experimentally the partial order  has been observed in $Gd_2Ti_2O_7$ \cite{POexp04}. Monte Carlo method has been utilized to study the nature of partial order in Ising model on  $kagom\acute{e}$ lattice \cite{Diep87}. There are exact results for the partially ordered systems which precede the above studies \cite{Diep87}-\cite{Diep04}.

The defining feature of the partial order is the profile of the magnetization-temperature curve. Below the N\'{e}el $T_N$ temperature the magnetization increases and has a local maximum at characteristic temperature $T^*$. Below $T^*$ the magnetization decreases and the zero temperature magnetization is equal to the difference between sublattice A and B zero temperature magnetization.

In the antiferromagnets, the onset of magnetism in sublattice A and B is at N\'{e}el temperature. When, during the preparation, a magnetic field is applied on sublattice's A and B electrons along the magnetic order of B electrons, the onset of magnetism in sublattice B is at $T_N$, while the onset of magnetism in sublattice A is at $T^*<T_N$. Within the interval $(T^*,T_N)$ the system is partially ordered, while below $T^*$ it is completely ordered. The effect of the applied magnetic field is absorbed into the characteristic temperature $T^*$ and sublattice magnetization $m$ (\ref{QCB5}) (which is effective value of the sublattice A saturated spin).

The novel result in the present paper is that a quantum partial order is realized at zero temperature.

\section {Magnon-induced superconductivity}

When the material is prepared with applied magnetic field $H=H_{cr2}$, the system is at quantum partial-order point (QPOP). The Zeeman splitting of sublattice $A$ electrons is zero and they do not contribute the spontaneous magnetization of the system. This is why we can write the Hamiltonian of the system at QPOP in terms of the fermion operators $c_{i\sigma }^ +,\, c_{i\sigma }$ (see Appendix A) and Holstein-Primakoff (HP) bose operators $a^+_j,\,a_j$ used to represent the spin operators of sublattice B localized electrons ${\bf S^B_j}(a^+_j,a_j)$. The Hamiltonian is a sum of three terms
\be\label{antiferro5} h=h^A+h^{AB}+h^B,\ee
where, at half-filling and accounting for the dispersions  (\ref{antiferro4c})
\bea\label{antiferro6}
 h^A & = & -t\sum\limits_{\ll ij \gg _A } {\left( {c_{i\sigma }^ + c_{j\sigma } + h.c.} \right)} \nonumber \\
 h^{AB} & = & \sqrt{\frac{s}{2}}J\sum\limits_{  \langle  ij  \rangle }\left(c_{i\downarrow }^ + c_{i\uparrow }a_j+c_{i\uparrow}^ + c_{i\downarrow }a_j^+\right) \\
h^B & = &  -  J_B\sum\limits_{  \ll  ij  \gg_B  } {{\bf S_i^B}
\cdot {\bf S_j^B}}\nonumber \eea
We represent the spin operators $\textbf{S}^B$, in the sublattice B Hamiltonian $h^B$, in terms of (HP) Bose operators $a^+_j,\,a_j$ keeping only the quadratic and quartic terms (see Appendix B). The next step is to represent the Hamiltonian $h^B$ in the Hartree-Fock approximation:
\be\label{antiferro7}
h^B\approx h^B_{HF} = \frac {J_B\, u}{2} \sum\limits_{  \ll  ij  \gg_B}(a_i^+a_i+a_j^+a_j-a_j^+a_i-a_i^+a_j) \ee
where $u$ is the Hartree-Fock parameter, to be determined self-consistently (see Appendix B). HF parameter depends on the temperature and
accumulates the result of the transversal spin fluctuations of the sublattice B localized spins. In the $h^{AB}$ part of the Hamiltonian (\ref{antiferro6}) the spin operators $\textbf{S}^B$ are approximated by linear terms of (HP) Bose operators $a^+_j,\,a_j$.
In momentum space representation Hamiltonian Eqs.(\ref{antiferro5}) adopts the form
\bea\label{antiferro7}
h & = & \sum\limits_{k\in B_r}\left[ \varepsilon_{k}^A c_{k \sigma}^+ c_{k \sigma}+\varepsilon_k^B a_k^+ a_k \right]  \\
& + & \frac {4J\sqrt{2s}}{\sqrt{N}}\sum\limits_{k q p \in B_r } \delta (p-q-k)\cos\frac {k_x}{2} \cos\frac {k_y}{2} \cos\frac {k_z}{2} \nonumber \\
& \times & \left(c_{p\downarrow }^ + c_{q\uparrow }a_k+c_{q\uparrow}^ + c_{p\downarrow }a_k^+\right),\nonumber \eea
with fermi $\varepsilon_k^A$ and bose $\varepsilon_{k \sigma}^B$ dispersions
\bea\label{antiferro8}
\varepsilon_{k}^A & = & -4t\left ( \cos k_x+\cos k_y+\cos k_z \right) \\
\varepsilon_k^B & = & 2J_B u \left ( 3-\cos k_x-\cos k_y-\cos k_z \right) \nonumber \eea
The two equivalent sublattices A and B of the body center cubic lattice are simple cubic lattices. Therefor the wave vectors $p,q,k$ run over the first Brillouin zone of a cubic lattice $B_r$ .

Let us average in the subspace of Bosons $(a^+,a)$-to integrate the Bosons in the path integral approach. In static approximation one obtains an effective fermion theory with Hamiltonian
\bea\label{antiferro9}
h_{eff} & = & \sum\limits_{k\in B_r}\varepsilon_{k}^A c_{k \sigma}^+ c_{k \sigma} - \frac 1N \sum\limits_{k_i p_i \in B_r} \delta (k_1-k_2-p_1+p_2) \nonumber \\
 & \times & V_{k_1-k_2} c_{k_1\downarrow}^+c_{k_2\uparrow}c_{p_2\uparrow}^+c_{p_1\downarrow} \eea
with potential
\be\label{antiferro10}
V_k = \frac {J^2 (1+\cos{k_x})(1+ \cos{k_y})(1+ \cos{k_z})}{J_B u \left ( 3-\cos k_x-\cos k_y-\cos k_z \right)} \ee

Following standard procedure one obtains the effective Hamiltonian in the Hartree-Fock approximation
\be\label{antiferro11}
h^{HF}_{eff}=  \sum\limits_{k\in B_r}\left[ \varepsilon_{k \sigma} c_{k \sigma}^+ c_{k \sigma}+\Delta_k c_{-k\downarrow}^+c_{k\uparrow}+\Delta_k^+c_{k\uparrow}c_{-k\downarrow}\right], \ee
with gap function
\be\label{antiferro12}
\Delta_k=\frac 1N \sum\limits_{p\in B_r}<c_{-p\uparrow}c_{p\downarrow}> V_{p-k} \ee
The Hamiltonian can be written in a diagonal form by means of Bogoliubov excitations $\alpha^+,\alpha,\beta^+,\beta$ with dispersions $E^{\alpha}_k=E^{\beta}_k=E_k$
\be\label{antiferro13}
E_k  = \sqrt{(\varepsilon_{k}^A)^2+|\Delta_k|^2}  \ee
In terms of the new excitations the gap equation reads
\bea\label{antiferro14}
\Delta_k= & - & \frac {1}{2N} \sum\limits_{p\in B_r}V_{k+p}\frac {\Delta_p}{\sqrt{(\varepsilon_{}^A)^2+|\Delta_p|^2}}\nonumber \\
& \times & \left(1-\frac {2}{e^{ {E_p}/{T}}+1}\right), \eea
where $T$ is the temperature.

\begin{figure}[!ht]
\epsfxsize=\linewidth
\epsfbox{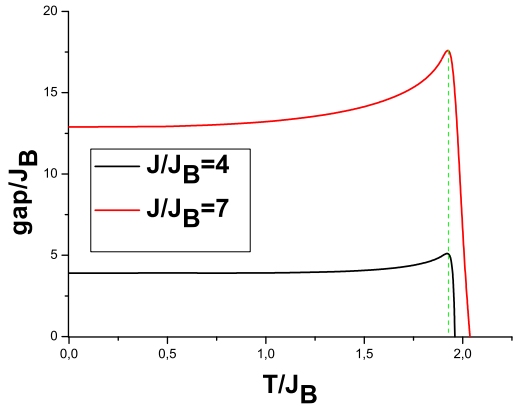} \caption{(Color online) The temperature dependence of the dimensionless gap $(gap/J_B)$ for $t/J_B=0.5$ and two different values of the parameter $J/J_B=7$ - upper (red) graph, $J/J_B=4$ - lower (black) graph. The vertical dash (green) line marks the N\'{e}el temperature.}\label{antiferro-gap-T}
\end{figure}

Having in mind that sublattices are simple cubic lattices and following the classifications for spin-triplet gap functions $\Delta_{-k}=-\Delta_{k}$ \cite{RKS10}, we obtained that the gap function with $T_{1u}$ configuration
\be\label{antiferr15}
\Delta_k=\Delta\left(\sin k_x+\sin k_y+\sin k_z \right) \ee
is a solution of the gap equation (\ref{antiferr15}) for some values of the temperature. The dimensionless gap ($gap/J_B$), as a function of dimensionless temperature ($T/J_B$), is depicted in Fig. (\ref{antiferro-gap-T}) for two different values of the parameter $J/J_B=4$, $J/J_B=7$ and $t/J_B=0.5$.

The parameters are chosen having in mind that $J$ is nearest neighbor exchange constant while $J_B$ next to nearest  neighbor exchange constant, therefore $J>J_B$. With value $J/J_B=7$ we overestimate the parameter to give better understanding of the phenomenon.

The figure shows that the temperature dependence of the gap  is quite unusual. The gap is approximately constant when the temperature is below the N\'{e}el temperature $T_N$, marked with vertical dash green line, weakly increases when the temperature approaches $T_N$ and abruptly falls to zero in paramagnetic phase. This is because the pairing of fermions, below the N\'{e}el temperature, is mediated by gapless bosons-magnons. The potential $V_k$  depends on temperature since the  Hartree-Fock parameter $u$ does.
Near the N\'{e}el temperature the  parameter $u$ decreases (Appendix B) and potential $V_k$ increases. Above N\'{e}el temperature the magnon opens a gap which rapidly increases when the temperature increases. This suppresses the superconductivity since the maximal value of the potential in paramagnetic phase is one over the magnon gap, so that when the magnon gap increases abruptly the potential decreases.

The term in (\ref{antiferro1}) which describes ferromagnetic Heisenberg exchange between sublattice A $(J_A>0)$ electrons can
contribute to the superconductivity through the Kohn-Luttinger mechanism. The results show\cite{RKS10} that the effect on the p-wave superconductivity with $T_{1u}$ configuration is weak. This permits us to drop it.

\section{Conclusion}

When, during the preparation of the material, magnetic field is applied along the sublattice B magnetization and switched off when the process is over there are two important consequences: i) Sublattice B electrons are localized, and the applied field do not affect sublattice B magnetization. There are experimental evidences for this. The magnetization-temperature \cite{spinelFeCr2S4} curve for FeCr$_2$S$_4$ spinel and magnetic susceptibility \cite{spinel+,spinel++} curves for MnV$_2$O$_4$ spinel increase below N\'{e}el $T_N$ temperature and have a local maximum at characteristic temperature $T^*$. The first part of the curves is not affected by the magnetic field.
At $T^*$ the system undergoes a partial order transition \cite{Diep97,Karchev08b,Karchev09b,Karchev15a}. Above $T^*$ only sublattice B electrons contribute to the magnetization, while below $T^*$ sublattice A and sublattice B electrons contribute to the magnetization of the system. This means that transversal fluctuations (magnons) of localized B electrons, which are responsible for magnetization above $T^*$, are long-range (gapless) in ZFC materials as well as in FC ones.
Therefore we have to study the spin fluctuations of sublattice B electrons by means of Heisenberg model without applied magnetic field (see Hamiltonian $h^B$ (\ref{antiferro6})).

 ii) When magnetic field is applied along the sublattice B magnetization, the magnetization-temperature curves for zero-field cooled  and field-cooled materials display a pronounced bifurcation below $T^*$. This is because the sublattice A and sublattice B electrons contribute to the magnetization of the system and sublattice A contribution decreases when the magnetic field, applied during the preparation of the material, increases. The explanation is usual. The Zeeman splitting of sublattice A electrons decreases when the magnetic field is applied. The nontrivial point is that it remains smaller and after the end of the process of the preparation of the material when the magnetic field is switched off. This is evident from the experimental magnetization-temperature graphs. To describe theoretically the phenomenon a theoretical magnetic field is included in the equations for the fermion dispersions (\ref{antiferro3}). It decreases the Zeeman splitting and the theoretical calculations of the magnetization-temperature dependence are in agreement with the experiment \cite{Karchev15a}. This is the only justification for introducing the theoretical field.

 In the present paper we use the Schwinger-bosons slave-fermions representation for the sublattice A fermions. At the same time we can use the original creation and annihilation Fermi operators. In the last case the Zeeman splitting determines the magnetization of the sublattice A electrons and the transversal fluctuation can be introduced by means of mean-field theory. This representation permits a smooth arrival at the Hamiltonian (\ref{antiferro6}), used to study superconductivity. The important point is that we can calculate the criteria for the insulator-metal transition (\ref{antiferro4}) and transition to superconductivity (\ref{antiferro41}) only approximately. The Schwinger-bosons slave-fermions representation represents the Coulomb term in quadratic form and one can calculate the contribution of the Coulomb repulsion exactly, which is very important especially to study insulator-metal transition. This representation do not permit a smooth arrival at the Hamiltonian (\ref{antiferro6}).

Hamiltonian written in terms of Schwinger-bosons and slave-fermions possesses $U(1)$ gauge invariance and one can introduce  $U(1)$ gauge fields.
The gauge field theory is used to describe the spin liquid phases of two dimensional antiferromagnets \cite{Sachdev89,Sachdev90} and the "pseudogap" phase of copper-based superconductors.
A key result \cite{Sachdev04,Vishvanath04} is that the gauge fields induced metallic states and free electron state are qualitatively different and cannot be adiabatically connected.
The basic aspects of the theory of gauge fields in insulator and metals are reviewed in \cite{Sachdev16}.

We consider 3D field-cooled spin-$1/2$ antiferromagnets. At temperature $T^*<T<T_{N}$ the state of the system is partially ordered. If we use Schwinger-bosons and slave-fermions representation for sublattice A electrons we can introduce $U(1)$ gauge fields which interact with transversal fluctuations (magnons) of sublattice B electrons. Because of sublattice B magnetization, the time-reversal symmetry is broken and the effective gauge theory, obtained integrating out the sublattice B magnons, has a term which breaks explicitly the time-reversal symmetry. The effective gauge theory is described by the Maxwell action together with a topological $\Theta$ term $S_{\Theta}=\frac {\Theta}{2\pi}\int d^3xdt\textbf{E}\cdot\textbf{B}$, where $\textbf{E}$ and $\textbf{B}$ are pseudo-electromagnetic fields and $\Theta$ is constant. In continuum limit, this term is total derivative. It is suggested in \cite{axion10}, that $\Theta$ becomes a dynamical pseudo scalar field associated with the magnetic fluctuations. The resulting system with slave fermions is the axion electrodynamics. Alternatively one can consider the original $U(1)$ gauge theory with topological term on a lattice. In both cases, the topologically nontrivial metal state differs from the metal state described  by the Hamiltonian (\ref{antiferro6}). They can not be adiabatically connected.

To finish,  it is important to underline that insulator-metal transition when the system is under hydrostatic pressure is a result of delocalization of all sublattice A and B electrons, while in the present case the applied magnetic field separates spatially electrons and only sublattice A ones are delocalized.

In the high-$T_c$ cuprates, $Fe$-based materials, two-dimensional organic compounds and heavy electron systems the superconductivity emergency is close to the magnetic quantum phase transition, where the magnetization of the system is weak or even zero \cite{Moriya03}.
With distinction in the present paper we have obtained a coexistence of superconductivity and magnetism with maximal magnetic moment. This is a novel result.

We focus on the bcc lattice structure of the material but the symmetry of superconducting order parameter is the same and for simple cubic lattice and for fcc lattice \cite{RKS10}.

\appendix
\section{}
{\bf Schwinger-Bosons Slave-Fermions Representation}
\vskip 0.51cm
We consider a theory with Hamiltonian
\be \label{QCB1}
\hat{h}^A  = -t\sum\limits_{\ll ij \gg _A } \left( c_{i\sigma }^ + c_{j\sigma}  + h.c. \right)
+ U \sum\limits_{i\in A} n_{i\uparrow} n_{i\downarrow} -\mu \sum\limits_{i\in A} n_i\ee

We represent the Fermi operators, the spin of the sublattice A electrons
\be\label{QCB1b}
S^{\nu A}_i=\frac 12 \sum\limits_{\sigma\sigma'}c^+_{i\sigma} \tau^{\nu}_{\sigma\sigma'}c^{\phantom +}_{i\sigma'},\ee where
$(\tau^x,\tau^y,\tau^z)$ are Pauli matrices, and the density operators $n_{i\sigma}$  in terms of the Schwinger bosons
($\varphi_{i,\sigma}, \varphi_{i,\sigma}^+$) and slave fermions
($h_i, h_i^+,d_i,d_i^+$). The Bose fields
are doublets $(\sigma=1,2)$ without charge, while fermions
are spinless with charges 1 ($d_i$) and -1 ($h_i$):
\begin{eqnarray}\label{QCB2} & & c_{i\uparrow} =
h_i^+\varphi _{i1}+ \varphi_{i2}^+ d_i, \qquad c_{i\downarrow} =
h_i^+ \varphi _{i2}- \varphi_{i1}^+ d_i, \nonumber
\\
& & n_i = 1 - h^+_i h_i +  d^+_i d_i,\quad  s^{\nu}_i=\frac 12
\sum\limits_{\sigma\sigma'} \varphi^+_{i\sigma}
{\tau}^{\nu}_{\sigma\sigma'} \varphi_{i\sigma'},\nonumber
\\& &
c_{i\uparrow }^+c_{i\uparrow }c_{i\downarrow }^+c_{i\downarrow}=d_i^+d_i \eea

\be\label{QCB2b}
\varphi_{i1}^+ \varphi_{i1}+ \varphi_{i2}^+ \varphi_{i2}+ d_i^+
d_i+h_i^+ h_i=1  \ee
To solve the constraint (Eq.\ref{QCB2b}), one makes a change of variables, introducing
Bose doublets $\zeta_{i\sigma}$ and
$\zeta^+_{i\sigma}\,$\cite{Schmeltzer}
\begin{eqnarray}\label{QCB3}
\zeta_{i\sigma} & = & \varphi_{i\sigma} \left(1-h^+_i h_i-d^+_i
d_i\right)^
{-\frac 12},\nonumber \\
\zeta^+_{i\sigma} & = & \varphi^+_{i\sigma} \left(1-h^+_i h_i-d^+_i
d_i\right)^ {-\frac 12},
\end{eqnarray}
where the new fields satisfy the constraint
$\zeta^+_{i\sigma}\zeta_{i\sigma}\,=\,1$. In terms of the new fields
the spin vectors of the sublattice A electrons Eq.(\ref{QCB1b}) have the form
\be
S^{\nu A}_{i}=\frac 12 \sum\limits_{\sigma\sigma'} \zeta^+_{i\sigma}
{\tau}^{\nu}_{\sigma\sigma'} \zeta_{i\sigma'} \left[1-h^+_i
h_i-d^+_i d_i\right] \label{QCB4} \ee
When, in the ground state,
the lattice site is empty, the operator identity $h^+_ih_i=1$ is
true. When the lattice site is doubly occupied, $d^+_id_i=1$. Hence,
when the lattice site is empty or doubly occupied the spin on this
site is zero. When the lattice site is neither empty nor doubly
occupied ($h^+_ih_i=d^+_id_i=0$)  $\,\,{\bf S}^A_{i}=1/2
{\bf n}_i,\,\,$ where the unit vector
\be\label{QCB5b}
n^{\nu}_i=\sum\limits_{\sigma\sigma'} \zeta^+_{i\sigma}
{\tau}^{\nu}_{\sigma\sigma'} \zeta_{i\sigma'}\qquad ({\bf
n}_i^2=1)\ee identifies the local orientation of the spin of the
sublattice A electron.

The Hamiltonian Eq.(\ref{QCB1}), rewritten in terms of Bose fields Eq.(\ref{QCB3}) and slave fermions, adopts the form
\bea\label{QCB3a}
\hat{h}^A  & = & -t\sum\limits_{\langle ij \rangle} \left[\left ( d^+_j d_i-h^+_j h_i \right) \zeta^+_{i\sigma}\zeta_{j\sigma}\right. \nonumber \\
& + & \left.\left ( d^+_j h^+_i-d^+_i h^+_j\right )\left (\zeta_{i1}\zeta_{j2}-\zeta_{i2}\zeta_{j1}\right ) + h.c. \right]\nonumber \\
& \times & \left(1-h^+_i h_i-d^+_id_i\right)^{\frac 12}\left(1-h^+_j h_j-d^+_jd_j\right)^{\frac 12} \nonumber \\
& + & U \sum\limits_i d^+_id_i -\mu \sum\limits_i \left (1-h^+_ih_i+d^+_id_i\right),\eea
To proceed we approximate the hopping term of the Hamiltonian Eq.(\ref{QCB3a}) setting  $\left(1-h^+_i h_i-d^+_id_i\right)^{\frac 12}\sim 1$ and keeping only the quadratic, with respect to fermions, terms.
Further, we represent the resulting Hamiltonian $h^A\approx h$ as a sum of two terms
\be\label{QCB4a}
h=h_0 + h_{int}, \ee
where
\bea\label{QCB4b}
h_0 = & - & t\sum\limits_{\langle ij \rangle} \left ( d^+_j d_i-h^+_j h_i + h.c.\right)
 +  U \sum\limits_i d^+_id_i \nonumber \\
& - & \mu \sum\limits_i \left (1-h^+_ih_i+d^+_id_i\right),\eea
is the Hamiltonian of the free $d$ and $h$ fermions, and
\bea\label{QCB4c}
h_{int} = & - &t\sum\limits_{\langle ij \rangle} \left[\left ( d^+_j d_i-h^+_j h_i \right) \left (\zeta^+_{i\sigma}\zeta_{j\sigma}-1\right)\right. \\
& + & \left.\left ( d^+_j h^+_i-d^+_i h^+_j\right )\left (\zeta_{i1}\zeta_{j2}-\zeta_{i2}\zeta_{j1}\right ) + h.c. \right]\nonumber\eea
is the Hamiltonian of boson-fermion interaction.

The ground state of the system, without accounting for the spin fluctuations, is determined by the free-fermion Hamiltonian $h_0$ and is labeled by the density of electrons
\be\label{QCB4d} n=1-<h^+_i h_i>+<d^+_id_i> \ee (see equation (\ref{QCB2})) and the zero temperature spontaneous dimensionless magnetization, of the sublattice A electron
\begin{equation}
m=\frac 12 \left(1-<h^+_i h_i>-<d^+_id_i>\right). \label{QCB51}
\end{equation}
At half-filling

\be\label{QCB4e} <h^+_i h_i>=<d^+_id_i>. \ee To solve this equation, for all values of the parameters $U$ and $t$, one sets the chemical potential $\mu=U/2$. Utilizing this representation of $\mu$ we calculate the magnetization "m" as a function of the ratio $t/U$.

In terms of the Schwinger bosons and slave fermions the mixed term in Hamiltonian (Eq.1) of the paper  adopts the form
\bea\label{supp1}
\hat{h}^{AB} & = &  J\sum\limits_{  \langle  ij  \rangle } {{\bf S_i^A}}\cdot {\bf S_j^B} \\
& = & J\sum\limits_{  \langle  ij  \rangle } {\frac 12 \left[1-h^+_i h_i-d^+_i d_i\right]}{\bf n_i}\cdot {\bf S_j^B} \nonumber
\eea
where ${\bf n}$ is the unit vector (Eq.\ref{QCB5b}). To proceed, one uses Holstein- Primakoff representation of the spin operators of sublattice B localized electrons ${\bf S}^B_j(a^+_j , a_j )$, where $a^+_j , a_j$ are Bose fields. In the free fermion approximation, when the interaction of fermions with $a^+_j , a_j$  Bose fields is not accounted for,  ${\bf S_j^B}=(0,0,s^B)$ and ${\bf n}_i=(0,0,-1)$, where antiferromagnetic order is accounted for. Hence, the contribution of the Hamiltonian (\ref{supp1}) to the free fermion Hamiltonian is
\be\label{supp2}
\hat{h}^{AB}\approx
 4Js^B\sum\limits_{  i\in A}  \left[h^+_i h_i+d^+_i d_i\right]\ee
with $s^B=1/2$ in our model.

Finally, we consider the term with applied, along sublattice B magnetization, magnetic field $H$
\be\label{supp3}
 \hat{h}^{H} =  - H \sum\limits_{  i \in A } S_i^{zA} \ee
In terms of the Schwinger bosons and slave fermions it has the form
\be\label{supp4}
 \hat{h}^{H} =  - H \sum\limits_{  i \in A } { {\frac 12 \left[1-h^+_i h_i-d^+_i d_i\right]} n^z_i}. \ee
Hence, the contribution to the free fermion Hamiltonian with  ${\bf n}_i=(0,0,-1)$ is
 \be\label{supp5}
 \hat{h}^{H}\approx -\frac {H}{2} \sum\limits_{  i \in A } \left[h^+_i h_i+d^+_i d_i\right]. \ee

 Collecting the above results one obtains the  Hamiltonian for the free $d$ and $h$ Fermions
 \be\label{sup6}
h_0 =   \sum\limits_{k\in B_r} \left (\varepsilon^d_k d_k^+ d_k + \varepsilon^h_k h_k^+ h_k \right)\ee
with dispersions
\bea\label{supp7}
\varepsilon^d_k & = & -4t\varepsilon_k +U-\mu+2J-\frac H2 \nonumber \\
\varepsilon^h_k & = & 4t\varepsilon_k+\mu+2J -\frac H2  \\
\varepsilon_k &  = & \left ( \cos k_x+\cos k_y+\cos k_z \right)\nonumber \eea
At half-filling $\mu=U/2$ and dispersions read
 \bea\label{supp7a}
\varepsilon^d_k & = & -4t\varepsilon_k +\frac U2+2J-\frac H2 \nonumber \\
\varepsilon^h_k & = & 4t\varepsilon_k+\frac U2+2J -\frac H2   \eea
The most important consequence is that the applied magnetic field compensates not only the Zeeman splitting
due to sublattice A and sublattice B spins exchange ($2J$ term), but and the Coulomb repulsion ($U$ term).
This is the grounding in understanding the novel mechanism of insulator-metal transition.

When the Zeeman splitting is zero there are no transversal fluctuations and one can set the Bose fields  $\zeta_{i1}=0$ and $\zeta_{i2}=1$. Then the representation (\ref{QCB2}) of the Fermi operators  adopts the form
\be\label{QCB22}
c_{i\uparrow} = d_i\sqrt{1 - h^+_i h_i - d^+_i d_i}, \qquad
c_{i\downarrow} = h_i^+ \sqrt{1 - h^+_i h_i - d^+_i d_i},\ee
where the equation (\ref{QCB3}) is accounted for. Having in mind the Fermi statistics of the operators we obtain the identity
\be\label{QCB23}
\sqrt{1 - h^+_i h_i - d^+_i d_i}=1-\frac 12 (h^+_i h_i + d^+_i d_i)-\frac 14 h^+_i h_i d^+_i d_i \ee
and the representation
\bea\label{QCD24}
c_{i\uparrow} =  d_i - \frac 12 d_i h^+_i h_i  \nonumber \\
c_{i\downarrow}  =  h_i^+ -\frac 12 h_i^+ d^+_i d_i.\eea
In the effective theory we keep only quadratic, with respect to the fermions, terms. This permit us to use the approximate representation for the fermi operators
\be\label{QCB25}
c_{i\uparrow} = d_i, \qquad
c_{i\downarrow} = h_i^+.\ee

\vskip 0.51cm

{\bf Hartree-Fock approximation in spin theory}

\vskip 0.51cm
To study the spin fluctuations of the localized sublattice B electrons we consider  Heisenberg Hamiltonian
\be\label{supp8}
\hat{h}^{B} =  -  J^B\sum\limits_{  \ll  ij  \gg_B  } {{\bf S_i^B}
\cdot {\bf S_j^B}}\ee
using  (HP) representation of the spin operators ${\bf S^B_j}(a^+_j,a_j)$.
In terms of the (HP) Bose operators and keeping only the quadratic and quartic terms, Hamiltonian (Eq.\ref{supp8}) adopts the form
\bea\label{supp9}
 \hat{h}^{B} & = & s^B J^B\sum\limits_{\ll ij \gg _B }\left( a^+_i a_i\,+\,a^+_j a_j\,-\,a^+_j a_i\,-\,a^+_i a_j\right) \nonumber \\
 & + & \frac {J^B}{4} \sum\limits_{\ll ij \gg _B }\left[a^+_i a^+_j( a_i-a_j)^2 + (a^+_i- a^+_j)^2  a_i a_j\right] \nonumber \\ \eea
and the terms without operators are dropped.
The next step is to represent the Hamiltonian in the Hartree-Fock  approximation:
\bea\label{supp9a}
 \hat{h}^{B}\approx h_{HF} & = & h_{cl}+h_q  \\
h_{HF} & = & \frac 32 N J^B (u-1)^2 + \sum\limits_{k\in B_r}\varepsilon^B_k\,a_k^+a_k, \nonumber \eea
where $s^B=1/2$ is used, $N = N_A = N_B$ is the number of sites on  sublattices, $\varepsilon^B_k$ is HP Bosons' dispersion
\be\label{supp10}
\varepsilon_k^B  =  2J^B u \left ( 3-\cos k_x-\cos k_y-\cos k_z \right) \ee
and $u$ is the Hartree-Fock parameter. To obtain the equation for the Hartree-Fock parameter we consider
the free energy of a system with Hamiltonian $h_{HF}$  (Eq.\ref{supp9a})
\be\label{supp11}
\mathcal{F}  =  \frac 32 N J^B (u-1)^2 + \frac {T}{N} \sum\limits_{k\in B_r} \ln\left(1-e^{-\varepsilon^B_k/T} \right).\ee
Then, the equation for the Hartree-Fock parameter is
\bea\label{supp12}& & \partial\mathcal{F}/\partial u  =  0,  \\
u & = & 1- \frac 23\frac 1N \sum\limits_{k\in B_r}\frac {\left ( 3-\cos k_x-\cos k_y-\cos k_z \right)}{e^{2J^B u \left ( 3-\cos k_x-\cos k_y-\cos k_z \right)/T}-1 } \nonumber \eea
The solution of equation (\ref{supp12}) for Hartree-Fock parameter $u$, as a function of dimensionless temperature $T/J^B$, is depicted in (Fig.\ref{supp-u-T}).
\begin{figure}[!t]
\epsfxsize=\linewidth
\epsfbox{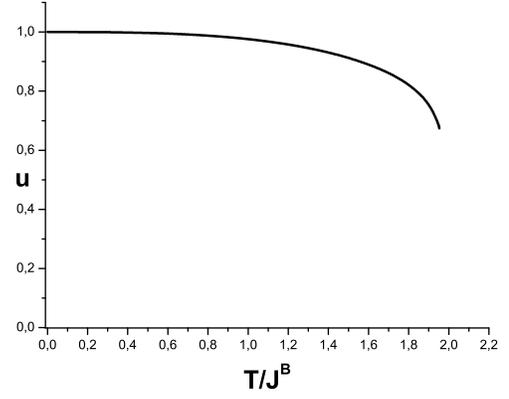}
\caption{ The temperature dependence of the Hartree-Fock parameter u as a function of dimensionless temperature $T/J^B$ for $s^B=1/2.$ }\label{supp-u-T}
\end{figure}
\begin{figure}[!t]
\epsfxsize=\linewidth
\epsfbox{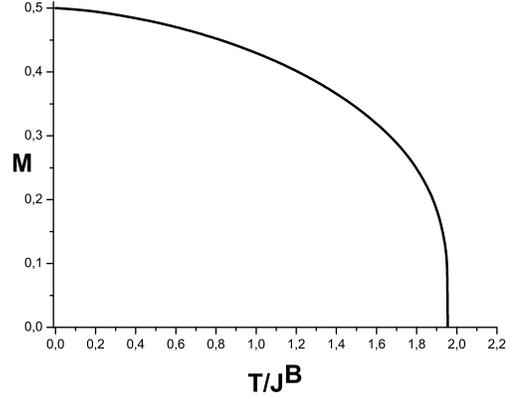}
\caption{ The temperature dependence of the spontaneous magnetization M as a function of dimensionless temperature $T/J^B$ for $s^B=1/2.$ }\label{supp-M-T}
\end{figure}

At quantum partial ordered point the spontaneous magnetization of the system equals the spontaneous magnetization of sublattice B localized electrons $M=M^B$,
\be\label{supp13}
 M = \frac 12 - \frac 1N \sum\limits_{k\in B_r}\frac {1}{e^{2J^B u \left ( 3-\cos k_x-\cos k_y-\cos k_z \right)/T}-1 }\ee
The spontaneous magnetization as a function of dimensionless temperature $T/J^B$ is depicted in (Fig.\ref{supp-M-T}). Figures (\ref{supp-u-T}) and (\ref{supp-M-T}) show that Hartree-Fock parameter $u$ decreases when the temperature approaches N\'{e}el temperature ($M=0$). This increases the potential which binds Cooper pairs of sublattice A electrons and the gap increases near the critical temperature.

Above N\'{e}el temperature we use Takahashi modified spin wave theory \cite{Takahashi86,Takahashi87}. The magnon opens a gap which increases rapidly
while Hartree-Fock parameter abruptly falls down to zero. As a result the binding potential is a constant which fast decrease is  reason for the suppression of superconductivity.

\end{document}